\documentclass[reprint, pra, amsmath,amssymb]{revtex4-1}
\usepackage{graphicx,amsmath,epsfig,latexsym}

\begin{document}

\author{Tyler D. Cumby}
\author{Ruth A. Shewmon}
\author{Ming-Guang Hu}
\author{John D. Perreault}
\author{Deborah S. Jin}
\affiliation{JILA, National Institute of Standards and Technology and Department of Physics, University of
Colorado, Boulder, Colorado  80309-0440, USA}

\title{Feshbach molecule formation in a Bose--Fermi mixture}
\date{\today}

\begin{abstract}
We investigate magnetoassociation of ultracold fermionic Feshbach molecules in a mixture of $^{40}$K and $^{87}$Rb atoms, where we can create $^{40}$K$^{87}$Rb molecules with a conversion efficiency as high as 45\%. In the limit of fast magnetic sweeps and small molecule fraction, we find that the timescale of molecule production is accurately predicted by a parameter-free model that uses only the atomic densities, masses, and the known Feshbach resonance parameters. We find that molecule production efficiency saturates for slower magnetic sweeps, at which point the maximum number of Feshbach molecules depends primarily on the atoms' phase-space density.  At higher temperatures, our measurements agree with a phenomenological model that successfully describes the formation of bosonic molecules from either Bose or Fermi gases.  However, for quantum degenerate atom gas mixtures, we measure significantly fewer molecules than this model predicts.
\end{abstract}

\maketitle

Weakly bound ultracold Feshbach molecules can be created in an atomic gas by sweeping a magnetic field across a Feshbach scattering resonance.  This process, called magnetoassociation, has proven to be one of the most robust methods of Feshbach molecule formation and is effective for both wide and narrow Feshbach resonances as well as for gases of fermionic atoms, bosonic atoms, or mixtures of two atomic species~\cite{Koehler2006, ChinRMP, Ferlaino2009}.  Magnetoassociation was an essential initial step in the recent success in creating a gas of ultracold polar molecules, where the Feshbach molecules were subsequently transferred to the ground ro-vibrational state~\cite{Ni2008}. This experiment starts from an ultracold Bose--Fermi gas mixture and creates fermionic $^{40}$K$^{87}$Rb molecules.  A quantum-degenerate Fermi gas of polar molecules would be an important step towards achieving novel quantum phases of matter~\cite{Krems2009}. However, the efficiency of the magnetoassociation is relatively low~\cite{Zirbel2008d} and this is a limiting factor in the number of ground-state molecules.  Fermionic Feshbach molecules have also recently been created in $^{23}\mathrm{Na}$-$^{40}\mathrm{K}$ and $^{6}\mathrm{Li}$-$^{23}\mathrm{Na}$ mixtures~\cite{Wu2012,Heo2012}. In all of these experiments, the fraction of atoms converted to molecules is less than 30\%.

%

In contrast, molecule conversion efficiencies as high as 90\% have been observed in homonuclear Fermi gases, where this process has been studied in detail ~\cite{Hodby2005}. The efficiency of magnetoassociation has been shown to depend on the magnetic-field sweep rate and on the phase-space density of the parent atomic gas. In the limit of sufficiently slow sweeps, a phenomenological model, which uses a pairing criteria based on the separation of two atoms in phase space and has a single fit parameter, was shown to describe the measured molecule creation efficiencies for different initial gas conditions~\cite{Hodby2005}.

Motivated by the relevance to ultracold polar molecule experiments as well as an interest in the nature of pairing in Bose--Fermi mixtures~\cite{Pieri2010,Zhang2012}, we report here an investigation of Feshbach molecule magnetoassociation in an ultracold gas mixture of $^{40}$K and $^{87}$Rb atoms.  Our measurements significantly extend the range of atom conditions studied for this mixture~\cite{Zirbel2008d, Klempt2008, Ospelkaus2006a, Olsen2009,Zirbel2008a}.  We characterize the density-dependent timescale of magnetoassociation as well as the roles of competing inelastic processes.

The remainder of this paper is organized as follows.  We briefly summarize our experimental methods in Sec.~\ref{scn:procedure}.  Section~\ref{scn:loss} describes the details of our magnetoassociation process with a particular focus on minimizing the effects of inelastic collisions.  In Sec.~\ref{scn:LZ} we show that the sweep-rate dependence of $^{40}$K$^{87}$Rb magnetoassociation follows a simple parameter-free model adapted from Landau--Zener physics. In Sec.~\ref{scn:saturation} we measure the saturated number of molecules that are produced under a wide range of atom conditions.  These are compared against the stochastic phase-space sampling model described in Ref.~\cite{Hodby2005}.

%

%

\section{Preparation of the Bose-Fermi mixture}
\label{scn:procedure}
	
Our experiments begin with an ultracold mixture of $^{87}$Rb atoms in the $|F,m_F\rangle=|1,1\rangle$ state and $^{40}$K atoms in the $|9/2,-9/2\rangle$ state, where $F$ is the total atomic angular momentum and $m_F$ is its projection onto the magnetic-field axis.  The gas  is confined in a far-detuned optical trap formed by a single beam of 1090 nm light focused to a 19 $\mu$m waist.  Typical radial and axial trapping frequencies for our measurements are 450 Hz and 6 Hz, respectively, for Rb \cite{Bnote}.  The K trap frequencies are higher than those of Rb by a factor of 1.4. We perform measurements above the phase transition temperature, $T_c$, for Bose--Einstein condensation (BEC) of the $^{87}$Rb gas, so that the spatial distributions of the two clouds are well-matched. The number of atoms in each species is varied between $5\times10^4$ and $6.5\times 10^5$, and the temperature, $T$, ranges from 250 nK to 770 nK.

The atomic samples are prepared at a magnetic field that is several Gauss above a broad s-wave Feshbach resonance between the $^{87}$Rb and $^{40}$K atoms.  Near the Feshbach resonance, the scattering length is given by $a = a_{bg}\left(1-\Delta/(B-B_0)\right)$, with $B_0=546.62$ G, $\Delta = -3.04$ G, and $a_{bg}=-187$ $a_o$, where $a_o$ is the Bohr radius~\cite{Klempt2008}.  We associate atoms into molecules by sweeping the magnetic field downwards across the resonance.  One pair of magnetic coils is used to provide the bulk of the bias field, while a much smaller pair produces fast magnetic-field changes on the order of 10 G via controlled discharge of a 30 mF capacitor.  With these coils, and employing open-loop corrections of transient magnetic fields caused by eddy currents~\cite{Olsen2009}, we can realize magnetic-field sweeps as fast as 300 G/ms. This allows us to explore the magnetic-sweep-rate dependence of molecule formation.

\section{Magnetoassociation in the presence of inelastic loss}
\label{scn:loss}
In the presence of inelastic collisions, the detected number of molecules can differ from the number that was created. For example, a molecule-atom collision can result in vibrational relaxation of the molecule prior to detection~\cite{Mukaiyama2004}. In addition, three-body inelastic collisions of atoms can lead to significant inelastic loss and heating as the magnetic field approaches a Feshbach resonance. This modifies the atom gas conditions relevant to magnetoassociation. To minimize these effects, we carefully adjust the initial atom gas conditions and our magnetoassociation procedure. This facilitates a comparison of our data to theories that ignore losses and heating.

\begin{figure}
 \includegraphics[width=0.5\textwidth]{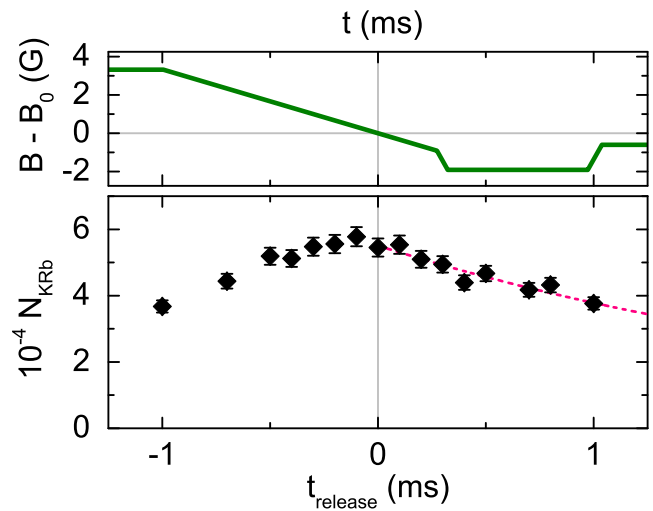}
 \caption{(Color online) Top: A typical magnetic-field sweep for molecule association.  Bottom: Measured molecule number vs the time at which the optical trap is suddenly turned off to release the gas, $t_{\rm{release}}$.  The error bars in all figures correspond to plus or minus the standard error obtained from repeated measurements. Here, $t=0$ is defined as the time when $B$ crosses the Feshbach resonance.  For increasing $t_{\rm{release}}>0$, the molecules spend more time in the trap where atom-molecule collisions can cause loss.  For comparison, the dashed line shows an exponential decay with a time constant of 2.7 ms.  For $t_{\rm{release}}<0$, the expansion of the atom gas prior to magnetoassociation eventually reduces the efficiency of molecule creation for a fixed magnetic-field sweep rate.  For this data, the initial atom gas has $3.6\times 10^5$ $^{40}$K atoms and $1.4\times 10^5$ $^{87}$Rb atoms at $T=380$ nK ($T/T_c=1.7$). }
 \label{fig:beta}
 \end{figure}

The top part of Fig.~\ref{fig:beta} illustrates a typical magnetoassociation sequence.  First, the field is swept at a rate of 3.33 G/ms from 549.9 G down to 545.7 G to form Feshbach molecules. To selectively image only the molecules, we  sweep the field in 50 $\mu s$ down to 544.7 G  where unpaired $^{40}$K atoms are transferred from the $|9/2,-9/2\rangle$ state to $|9/2,-7/2\rangle$ state using RF Adiabatic Rapid Passage (ARP).  At this magnetic field, the binding energy of the molecules is $h\times$3 MHz~\cite{Zirbel2008d}, where $h$ is Planck's constant, and the ARP for atoms does not affect the molecules.  To image the molecules, we then sweep the field in 65 $\mu s$ back up to 546.0 G, where the molecules can be selectively imaged with light tuned to the cycling transition of $^{40}$K. We find that the molecule number measured with this imaging technique is 85$\pm$5\% of the number measured after dissociating the molecules (by sweeping the magnetic field upwards across the Feshbach resonance) and imaging the resulting atoms.  Therefore, when imaging the molecules below the Feshbach resonance, we apply a multiplicative factor to correct the number for this measurement inefficiency.

To minimize the loss of molecules due to inelastic collisions, we can release the gas from the optical trap after the molecules are formed. The sudden turn-off of the optical trap initiates an expansion of the gas that rapidly lowers the densities and switches off inelastic collisions.   Figure~\ref{fig:beta} shows the measured molecule number as a function of the time that the optical trap is turned off, $t_{\rm{release}}$, where $t=0$ is defined as the time when the magnetic field crosses the Feshbach resonance.  For $t_{\rm{release}}>0$ we observe a decrease in the molecule number with increasing $t_{\rm{release}}$. We attribute this loss to inelastic collisions as the molecules spend more time in a relatively high density atom-molecule gas mixture in the trap.  For $t_{\rm{release}}<0$, the magnetic field crosses the Feshbach resonance after the gas starts expanding.  For short expansion times, where $-0.3<t_{\rm{release}}<0$, the measured number of molecules is constant within the measurement uncertainty.  For longer expansion times, corresponding to $t_{\rm{release}}<-0.3$, the rapidly dropping atom gas density reduces the efficiency of magnetoassociation for the 3.33 G/ms sweep rate.  Thus, we find $t_{\rm{release}}=0$ gives the maximum number of molecules.

\begin{figure}
 \includegraphics[width=0.5\textwidth]{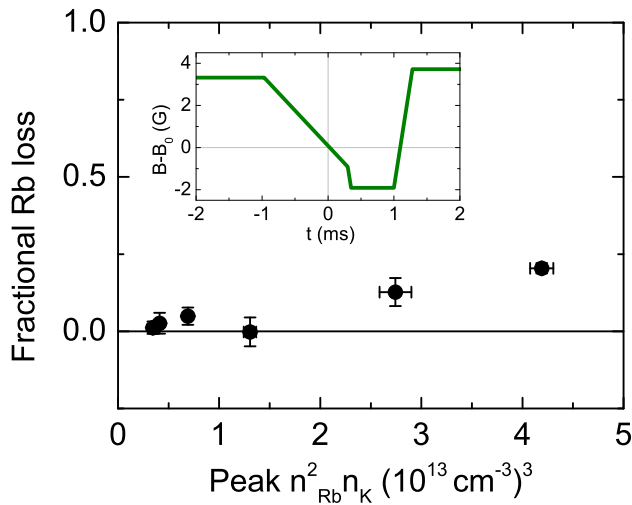}
 \caption{(Color online) Fractional loss of the $^{87}$Rb atoms after a double magnetic-field sweep to create and then dissociate $^{40}$K$^{87}$Rb Feshbach molecules. The measured loss is plotted as a function of the product of the initial peak $^{40}$K density and the peak $^{87}$Rb density squared. Here, the initial atom gas has $3\times 10^5$ $^{87}$Rb atoms and $2\times 10^5$ $^{40}$K atoms with a Rb radial (axial) trap frequency of 490 (6) Hz. The atom densities were varied by modulating the optical trap intensity to parametrically heat the cloud to temperatures between 490 nK and 730 nK. Inset: magnetic double-sweep sequence.  The optical trap is turned off at $t$=0 as the magnetic field crosses the Feshbach resonance.}
 \label{fig:alpha}
 \end{figure}

Another issue is that inelastic collisions can cause loss and heating of the atoms even before molecules are formed.  In particular, three-body recombination is an inelastic collision process for three atoms that produces a molecule plus an atom, both with excess kinetic energy.  The rate for three-body recombination increases dramatically near the Feshbach resonance.  To estimate the size of this effect, we dissociate the molecules by sweeping the magnetic field back up across the Feshbach resonance at a rate of 20 G/ms. Since the adiabatic magnetoassociation process is reversible, this sequence should conserve the atom number in the absence of inelastic collisions. Figure~\ref{fig:alpha} shows the measured fractional loss of $^{87}$Rb atoms as a function of the initial gas density.  For a Bose--Fermi mixture, the dominant three-body recombination process removes two bosons and one fermion from the trap. Characterizing the initial atom conditions by the peak $^{87}$Rb density squared times the peak $^{40}$K density,  we find that the atom loss is below 10\% as long as this density product remains lower than 2$\times(10^{13}$ cm$^{-3})^3$.  The initial atom conditions for data in Sections~\ref{scn:LZ} and~\ref{scn:saturation} were chosen to satisfy this condition.

Although lower density conditions were chosen to characterize the molecule formation process, we note that our largest molecule clouds are produced from higher atom densities.  For example, we created $7\times 10^4$ $^{40}$K$^{87}$Rb molecules from a mixture whose density product $n_{\text{Rb}}^2 n_\text{K}$ was  $3\times(10^{13}$ cm$^{-3})^3$.  The initial gas in this case had $6.5\times 10^5$ $^{40}$K atoms and $2.1\times 10^5$ $^{87}$Rb atoms at 560 nK, where $T/T_F$ = 0.51 and $T/T_c$ = 2.0.

\section{Dependence of magnetoassociation on the magnetic-field sweep rate}
\label{scn:LZ}

\begin{figure}
 \includegraphics[width=0.5\textwidth]{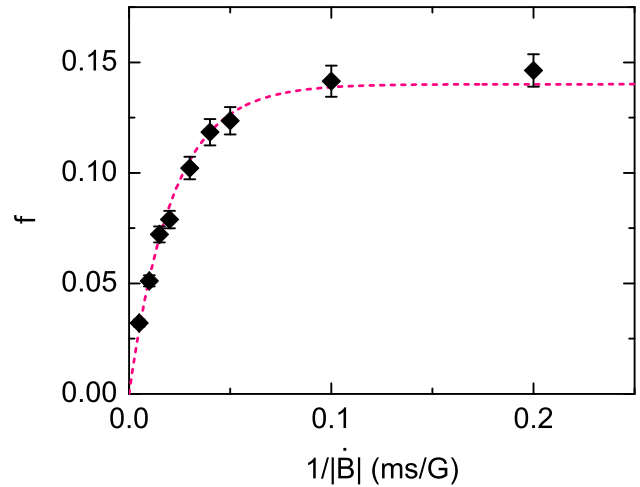}
 \caption{(Color online) Fraction of the minority atom species converted to molecules, $f$,  vs the inverse magnetic-sweep rate, $1/|\dot{B}|$, of a 7 G linear ramp of the magnetic field across the Feshbach resonance.  The dashed line shows a fit to Eq.~(\ref{eqn:generalizedLZ}), which gives $\Gamma=6.5(5)$ G/ms and $f_0=14(2)\%$, or $3.8(5)\times10^4$ molecules, for this data. The initial gas consisted of $2.6\times 10^5$ $^{40}$K atoms and $3.4\times10^5$ $^{87}$Rb atoms at $T$= 490 nK.  }
 \label{fig:LZcurve}
 \end{figure}

The efficiency of molecule creation, $f$, which is defined here as the number of molecules divided by the lesser of the initial numbers of atoms of $^{40}$K and $^{87}$Rb, $N_< $, depends on the magnetic-field sweep rate as shown in Fig.~\ref{fig:LZcurve}.   Only a small number of molecules are formed when the field sweeps too quickly across the Feshbach resonance.  As the sweep becomes slower and therefore more adiabatic, molecule production increases and finally saturates. The saturated conversion efficiency for Bose--Fermi mixtures is typically well below unity, as is the case in Fig.~\ref{fig:LZcurve}. In this section, we focus on the fast-sweep regime, where a simple extension of a two-body picture predicts that the molecule number increases linearly with the sweep duration~\cite{Chwedenczuk2004, Koehler2006}.   The maximum molecule conversion efficiency in the saturated regime is the topic of Sec.~\ref{scn:saturation}.

In order to extract information about the fast-sweep and saturated regimes independently, we fit the molecule conversion efficiency as a function of the magnetic-sweep rate $|\dot{B}|$ to the following formula:
 \begin{equation}
 f = f_0\left( 1-e^{-\Gamma/(f_0|\dot{B}|)}\right).
 \label{eqn:generalizedLZ}
 \end{equation}
Here, $f_0$ is the saturated molecule conversion fraction and $f=\Gamma/|\dot{B}|$ in the fast-sweep limit.

 In the fast-sweep regime, the molecule number is predicted to follow a Landau--Zener-like behavior~\cite{Zener1932, Koehler2006} with the transition probability into the molecular state given by $P = 1 - e^{-2\pi\delta_{LZ}} \approx   2\pi\delta_{LZ}$.  For two particles in a box of volume $\mathcal{V}$, the Landau--Zener parameter $\delta_{LZ}$ depends on the Feshbach resonance parameters and is given by~\cite{Chwedenczuk2004}
\begin{equation}
 \delta_{LZ} = \frac{1}{\mathcal{V}}\frac{2 \pi \hbar}{ \mu}\left| \frac{a_{bg} \Delta}{\dot{B}} \right|
 \label{eqn:dLZ0}
 \end{equation}
where $\hbar=h/2 \pi$ and $\mu$ is the two-body reduced mass. Interestingly, this result applies for thermal atoms (fermions or bosons) as well as for atoms in the motional ground-state, such as atoms in a Bose--Einstein condensate~\cite{Chwedenczuk2004}. Using classical probability theory, Eq.~(\ref{eqn:dLZ0}) can be generalized to multiple particles simply by multiplying $P$ by the total number of available atom pairs to get the number of molecules~\cite{Chwedenczuk2004}.  Using a local density approximation, the molecule density at position ${\bf r}$ is then given by
 \begin{equation}
n_{{\rm mol}}({\bf r})= n_1({\bf r})n_2({\bf r}){2 \pi} \frac{2 \pi \hbar}{\mu}\left| \frac{a_{bg} \Delta}{\dot{B}} \right|.
 \label{eqn:dLZ}
 \end{equation}
Integrating Eq.~(\ref{eqn:dLZ}) over the trapped atom gas distribution, and dividing by $N_<$, gives
 \begin{equation}
f= \langle n_>\rangle {2 \pi} \frac{2 \pi \hbar}{\mu}\left| \frac{a_{bg} \Delta}{\dot{B}} \right|.
 \end{equation}
where $\langle n_>\rangle = \frac{1}{N_<}\int n_1({\bf r})n_2({\bf r}) \mathrm{d}^3r$ describes the density overlap of the two atom clouds.

 \begin{figure}
 \includegraphics[width=0.5\textwidth]{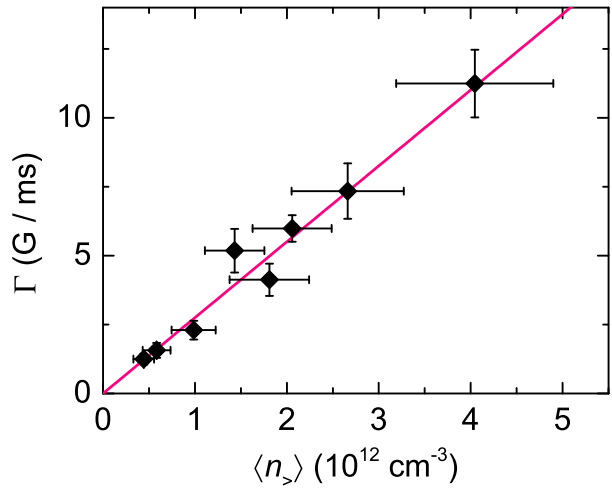}
 \caption{(Color online) The initial molecule conversion efficiency divided by the magnetic-field sweep rate, $\Gamma$, vs the density overlap of the two atom clouds.
The prediction for $\Gamma$ using the $^{40}$K-$^{87}$Rb Feshbach resonance parameters (Eq.~(\ref{eqn:Gvalue})) is shown as a solid line.  Here, the atomic densities are varied by associating molecules at different times during expansion from the optical trap (up to 1.2 ms), as well as by adjusting the evaporation trajectory and optical trap frequencies.  For this data, the $^{87}$Rb radial (axial) trap frequency ranged from 360 to 520 Hz  (4 to 7 Hz).}
 \label{fig:LZparameter}
 \end{figure}

Plugging the relevant Feshbach resonance parameters into Eq.~(\ref{eqn:dLZ}) yields
 \begin{equation}
 \Gamma/ \langle n_>\rangle = (2.8\pm0.1)\times10^{-12}~\mathrm{cm}^3~\mathrm{G}/\mathrm{ms}.
 \label{eqn:Gvalue}
 \end{equation}
This prediction for $\Gamma$, which is shown as a solid line in Fig.~\ref{fig:LZparameter}, agrees well with our measurements of $\Gamma$ as a function of the density overlap of the initial atom gas, $\langle n_>\rangle$.
From this we conclude that magnetoassociation in the perturbative regime of fast magnetic-field sweeps behaves as expected in our Bose--Fermi mixture. In the next section, we turn our attention to the saturated regime, which is important for experiments where one would like to create larger numbers of Feshbach molecules.

\section{Saturated molecule number}
\label{scn:saturation}

In the saturated regime of slow magnetic-field sweeps, molecule creation has been shown to depend on the entropy or phase-space density of the atom gas~\cite{Hodby2005,Williams2006,Watabe2008}.  A corresponding phenomenological model has successfully described molecule creation efficiency for homonuclear molecules in $^{40}$K and $^{85}\mathrm{Rb}$ gases as well as in  heteronuclear mixtures $^{85}\mathrm{Rb}$-$^{87}\mathrm{Rb}$ \cite{Papp2006}, $^{40}\mathrm{K}$-$^{87}\mathrm{Rb}$ \cite{Zirbel2008d}, and $^{6}\mathrm{Li}$-$^{40}\mathrm{K}$ \cite{Spiegelhalder2010}.  While previous experimental work with $^{40}\mathrm{K}$-$^{87}\mathrm{Rb}$ trapped gas mixtures focused on RF association~\cite{Zirbel2008d,Klempt2008}, we consider here magnetoassociation.

\begin{figure}
 \includegraphics[width=0.5\textwidth]{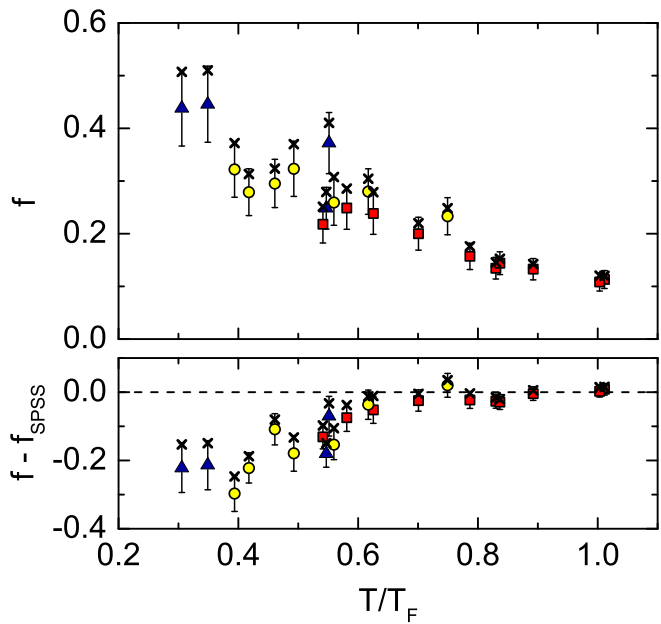}
 \caption{(Color online) Top: Measured conversion efficiency, $f$, vs the initial $T/T_F$ of the $^{40}$K atoms. The atom number ratio ranges from $1.6 \le N_K/N_{Rb} \le2$ (squares), $2 \leq N_K/N_{Rb}\le 4$ (circles), or $4 \le N_K/N_{Rb} \le7$ (triangles).  The crosses include a correction to $f$, based on an estimate of inelastic collisions between $^{40}$K$^{87}$Rb molecules and $^{87}$Rb atoms. The estimated shift of $f$ is between 0.01 and 0.07. Molecule conversion efficiency increases for higher atomic phase-space density (lower $T/T_F$). Bottom: The difference between the SPSS model and the data, $f-f_{SPSS}$, vs the initial $T/T_F$ of the $^{40}$K atoms.}
 \label{fig:Nmol}
 \end{figure}

Figure ~\ref{fig:Nmol} shows the measured molecule creation efficiency $f$ as a function of $T/T_F$ of the $^{40}$K gas, where $T_F$ is the Fermi temperature. Here, we varied the  magnetic-field sweep rate, using the findings of Sec.~\ref{scn:LZ} to ensure that the sweeps were sufficiently slow that the molecule number was in the saturated regime.  For this data, the number of $^{40}$K atoms is larger than the number of $^{87}$Rb atoms. Rb radial (axial) trap frequencies range from 360 to 550 Hz  (4 to 6 Hz), and initial atom temperatures are between 250 nK and 770 nK.   As was observed for homonuclear molecules~\cite{Hodby2005}, we find that the molecule conversion efficiency increases for higher initial atomic phase-space densities.

We compare our results with the Stochastic Phase-Space Sampling (SPSS) model that has been used previously to treat homonuclear molecule formation~\cite{Hodby2005}. This semi-classical model is based on the idea that two atoms are able to pair if they lie within a volume of relative phase-space set by an adjustable parameter $\gamma$:
 \begin{equation}
 \mu\left| v_{rel}\right| \left| r_{rel}\right| < \frac{h}{2}\gamma
 \label{eqn:criterion}
 \end{equation}
where $v_{\text{rel}}$ is the relative speed of the two atoms and $r_{\text{rel}}$ is their separation.  Another property of the model is that atoms are only allowed to pair once (i.e., paired atoms are removed from further consideration).  To determine the parameter $\gamma$ in our system, we fit the molecule creation efficiency data from Fig.~\ref{fig:Nmol} to the SPSS model.  We first use a Monte-Carlo calculation to generate atom distributions matching the initial conditions for each of our measurements. These atoms are then paired according to the criterion in Eq.~(\ref{eqn:criterion}) to determine the molecule creation efficiency.  A fit to the data with initial $T/T_F > 0.55$ returns a value of $\gamma_{BF}$=0.38(3). The error bar is dominated by a 10\% systematic uncertainty in the measured conversion fraction.  Our measured $\gamma_{BF}$ for $^{40}$K-$^{87}$Rb is consistent with the values $\gamma_F=0.38(4)$ found in $^{40}$K and $\gamma_B=0.44(3)$ measured in $^{85}$Rb~\cite{Hodby2005}.

In the bottom panel of Fig.~\ref{fig:Nmol}, we show that the SPSS model with $\gamma_{BF}$=0.38 significantly overestimates the conversion efficiency for atom clouds with low $T/T_F$.  We note that previous measurements for homonuclear Feshbach molecules in $^{40}$K agreed with SPSS predictions for $T/T_F$ as low as 0.05 \cite{Hodby2005}. The failure of the SPSS model to describe our measurements in the quantum-degenerate regime may be due to heating by inelastic collisions, or may suggest that this model is inadequate for describing the production of \textit{fermionic} molecules from quantum-degenerate atom gas mixtures.

The maximum molecule conversion efficiency we observe in Fig.~\ref{fig:Nmol} is 45\%, which occurs at our lowest $T/T_F$ for the initial $^{40}$K gas.  Starting from even lower $T/T_F$, or higher phase space density, would seem then to be an obvious way to increase the molecule conversion efficiency.  However, going to  higher initial phase space density in the Bose--Fermi mixture would require creating a BEC of $^{87}$Rb or having an even larger ratio of the number of fermionic atoms to bosonic atoms.  In either case, this would result in a reduction in the spatial overlap of the Bose gas and the Fermi gas in the trap, as well as in momentum space, which will presumably decrease the molecule conversion efficiency.  This effect has been observed below the BEC transition temperature $T_c$ for heteronuclear mixtures~\cite{Zirbel2008d, Papp2006}.  In addition, the high density of the BEC can result in a rapid loss of molecules due to inelastic atom-molecule collisions.

\begin{figure}
\includegraphics[width=0.5\textwidth]{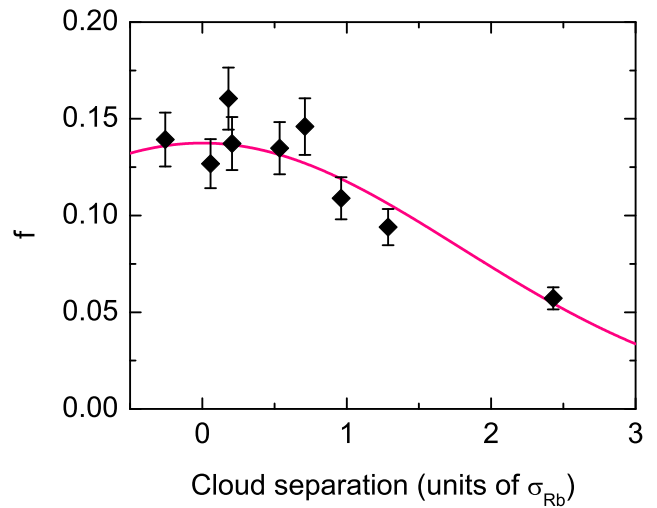}
\caption{(Color online) Molecule conversion efficiency, $f$, as a function of the separation between the $^{40}$K and $^{87}$Rb clouds in units of the width, $\sigma_\text{Rb}$,  of the $^{87}$Rb cloud, which is the smaller of the two atom clouds.  The solid curve shows an empirical fit to a Gaussian with an RMS width of 1.8(1) $\sigma_\text{Rb}$. For this data, molecules are created from $2.8\times10^5$ $^{87}$Rb atoms and $2.1\times 10^5$ $^{40}$K atoms at $T$ = 490 nK and $T/T_c$=1.6.}
\label{fig:separation}
\end{figure}

Another challenge for heteronuclear molecules, such as fermionic Feshbach molecules, is that differences in the masses, polarizabilities, and/or magnetic moments of the initial atoms can cause the two atom clouds to have different equilibrium positions in the trap.  To measure the significance of this effect on the molecule conversion efficiency, we intentionally separated the $^{40}$K and $^{87}$Rb clouds along the weak axial direction of the optical trap by applying a magnetic-field gradient before molecule formation.  The results of this measurement are shown in Fig.~\ref{fig:separation}, where we plot the saturated molecule conversion efficiency against the spatial offset of the two atom clouds.  We find that the conversion efficiency drops when the spatial separation is significant compared to the root-mean-squared (RMS) size of the $^{87}$Rb gas in the trap, $\sigma_\text{Rb}$. For the data in Fig.~\ref{fig:Nmol}, the spatial offset of the atom clouds in the axial direction is less than 1.2 $\sigma_\text{Rb}$ and the calculated offset in the vertical direction due to differential gravitational sag is less than 0.3 $\sigma_\text{Rb}$.  These relative displacements of the atom clouds were accounted for in the SPSS modeling.

\section{Conclusions}

We have studied the efficiency of converting atoms to fermionic Feshbach molecules using magnetic sweeps in a regime where inelastic collision effects are minimized. We find that in the regime of fast magnetic-field sweeps, the molecule creation efficiency is proportional to the inverse sweep speed with a coefficient that depends only on the Feshbach resonance parameters and the density overlap of the two atom clouds. For sufficiently slow magnetic sweeps, where the molecule number saturates, we find that the maximum conversion efficiency increases with increasing phase-space density. Our measurements with $T>T_c$ for $^{87}$Rb and $T/T_F>0.5$ for $^{40}$K agree with a phenomenological model having a single fit parameter.  However, significant discrepancies are observed between this model and our data for a more quantum-degenerate Fermi gas. The highest conversion efficiency, which is about 45\% for our data, is achieved for larger numbers of fermionic atoms than bosonic atoms and for $T$ just above $T_c$.

\acknowledgements{The authors thank Chen Zhang and Chris Greene for useful discussions.  This work was supported by NIST, the NSF, and NDSEG.}

\end{document}